\documentstyle[psfig]{l-aa}
\begin{document}
\def\fu{$f_1$}
\def\t{$\pm$}
\def\fd{$f_2$}
\def\fdu{$f_2 - 2f_1$}
\def\fp{$f_1 + f_2$}
\def\fm{$f_2 - f_1$}
\def\cd{cd$^{-1}$}
\def\cds{cd$^{-1}$\,}
\def\kms{km~s$^{-1}$}
\def\kmss{km~s$^{-1}$\,}
\def\I{\'\i}
\def\salp{\vskip 0.3truecm}
\thesaurus{6(03.13.2 - 08.15.1 - 08.22.1 - 10.19.2)}
\title{The galactic double--mode Cepheids}
\subtitle{II. Properties of the generalized phase differences}
\author{E. Poretti\inst{1} and I. Pardo\inst{2}}
\institute {Osservatorio Astronomico di Brera, Via Bianchi 46,
I-22055 Merate, Italy\\E--mail: poretti@merate.mi.astro.it
\and
Universit\`a di Milano, I-20100 Milano, Italy}
\offprints{E. Poretti}
\date{Received date; Accepted Date}
\maketitle
\markboth{E. Poretti and I. Pardo: The galactic double--mode Cepheids
II. The Fourier parameters}{ }
\begin{abstract}
By considering the least--squares fits of the double--mode Cepheid light curves
discussed in Paper I we defined their properties by their Fourier
parameters and generalized phase differences $G_{i,j}$. When plotting the latter
quantities as a function of the respective order, the second order terms 
are confined in the region just below 3/2$\pi$;
the third order terms have $\pi/2 < G_{i,j} < \pi$, the fourth order ones
cluster around 2$\pi$ (or 0), the fifth order ones seem to have
$\pi < G_{i,j} < 3/2 \pi$. The mean $G_{i,j}$ values are also regularly spaced.
The progression of the $G_{i,j}$ values as a function of the period was
investigated and the signature of a possible resonance near 6.0 d was found.
\end{abstract}

\section{Introduction}
Pardo \& Poretti (1996, hereinafter Paper I) made a frequency analysis of
the available photometry of galactic double--mode Cepheids (DMCs) and obtained
a very reliable set of Fourier parameters for each star. However,
a fully comprehensive and synthetic observational
description of the DMC light curves is still lacking. In this work, 
directly originating from the previous one, we try to determine  the
common characteristics of the DMC light curves, searching for the boundary
values of the Fourier parameters and regularities in their progression.

\section{The generalized phase differences}
In Paper I  we fitted the $V$ magnitudes  by means of the formula
\begin{equation}
V(t)= V_o + \sum_z {A_z \cos [2\pi f_z  (t-T_o) +\phi_z ]}
\end{equation}
where $f_z$ is the generic frequency, which can be an independent 
frequency (\fu and \fd), a harmonic or a cross coupling term. 
Our previous analysis demonstrated that each component
in the DMC light curves  can be defined as a combination of 
two basic frequencies \fu and \fd; by defining $z=(i,j)$, we have
$f_z=f_{i,j}$=$i$\fu+$j$\fd. Some examples:
for $(i,j)$=2,0 we have the harmonic 2\fu; for $(i,j)$=1,1 the \fp term;
for $(i,j)$=--1,1 the \fm term; for $(i,j)$=3,2 the 3\fu--\fd and so on.

In order to define the properties of the Fourier parameters of the DMC light
curves it is very useful to recall to mind the {\it generalized phase differences}
introduced by Antonello (1994b A\&A 291, 820), here noted as $G_{i,j}$.
They are a linear combination of the phases of
each term $f_{i,j}$ and of the phases $\Phi_1$ and $\Phi_2$ of the
independent frequencies \fu and \fd ($\Omega_1$ and $\Omega_2$ in Antonello's
notation). Their expression is given by 
%\begin{equation}
%G_z=\phi_z - i\Phi_1 - j\Phi_2 + 2k\pi
%\end{equation}
%or
\begin{equation}
G_{i,j}=\phi_{i,j} - i\Phi_1 - j\Phi_2 + 2k\pi
\end{equation}

The numerical application to the TU Cas fit provides some examples (the 
integer $k$ values have to be selected so that $G_{i,j}\in[0,2\pi]$):
\[ G_{1,1}=\phi_{1,1} - \Phi_1 - \Phi_2 + 2k\pi = \]
\[ \hspace*{1.0truecm} = 4.13 - 4.31 - 2.19 + 2\pi = 3.91\]
\[ G_{-1,1}=\phi_{-1,1} + \Phi_1 - \Phi_2 + 2k\pi = \]
\[ \hspace*{1.0truecm} = 1.92 + 4.31 - 2.19 = 4.04 \]
\[ G_{3,2}=\phi_{3,2} - 3\Phi_1 - 2\Phi_2 + 2k\pi = \]
\[ \hspace*{1.0truecm} = 2.11 - 3\cdot 4.31 -  2\cdot 2.19 + 6\pi = 3.64 \]
\begin{table*}
\begin{flushleft}
\caption{Generalized phase differences for all the galactic DMCs;
their measure units are
in rad 10$^{-2}$.  The amplitude ratios $R_{21}$ for the \fu and \fd 
frequencies are also reported. Except for the unique 1$O$/2$O$ pulsator
CO Aur, the stars are listed in order of increasing period}
\begin{tabular}{l  lll  llll}
\hline
\hline
\noalign{\smallskip}
 & \multicolumn{1}{c}{TU Cas} & \multicolumn{1}{c}{U TrA} & \multicolumn{1}{c}{VX Pup} &
   \multicolumn{1}{c}{AS Cas} & \multicolumn{1}{c}{AP Vel} & \multicolumn{1}{c}{BK Cen} &
   \multicolumn{1}{c}{UZ Cen}\\
\noalign{\smallskip}
\hline
\noalign{\smallskip}
R$_{21}$(\fu)&346\t4 & 323\t4 & 184\t6 & 305\t10 & 283\t4 & 263\t8 & 326\t11 \\
R$_{21}$(\fd)&123\t9 & 99\t20 & 118\t7 & 139\t15 & 117\t15 & 102\t19 &122\t37\\
\noalign{\smallskip}
2\fu    &  415\t3 &  415\t3 & 414\t8  &  415\t5 & 417\t4  &  423\t5  & 420\t6 \\
\fp     &  391\t4 &  404\t3 & 414\t5  &  412\t4 & 396\t5  &  427\t7  & 416\t11\\
\fm     &  404\t6 &  374\t3 & 450\t9  &  409\t7 & 441\t6  &  445\t11 & 393\t20\\
2\fd    &  433\t10& 439\t10 & 447\t13 &  434\t11& 445\t13 &  488\t23 & 449\t39\\
\noalign{\smallskip}
3\fu    &  207\t6 &   208\t6& 231\t33 & 219\t15 & 231\t9  &  208\t11 & 216\t10\\
2\fu+\fd&  189\t6 &   246\t5& 206\t13 & 210\t8  & 222\t9  &  214\t9  & 226\t14\\
\fu+2\fd &  178\t11&  181\t9& 228\t17 & 243\t12 & 202\t17 &  230\t23 & \\
2\fd--\fu &  165\t21&       & 238\t25 & 219\t18  \\
\noalign{\smallskip}
4\fu    &  604\t13&  588\t13&         &         &         &          & 642\t26\\
3\fu+\fd&  608\t9 &  623\t8 & 668\t34 & 629\t19 & 588\t27 &  621\t24 & 634\t21\\
2\fu+2\fd& 586\t13 \\
3\fu--\fd&         &        &         &         &         &          & 687\t52\\
\noalign{\smallskip}
3\fu+2\fd& 364\t21  \\
4\fu+\fd&  386\t15&  407\t15 \\
\noalign{\smallskip}
\hline
\hline
\noalign{\smallskip}
   & \multicolumn{1}{c}{Y Car} & \multicolumn{1}{c}{AX Vel} &
   \multicolumn{1}{c}{GZ Car} & \multicolumn{1}{c}{BQ Ser} & \multicolumn{1}{c}{EW Sct} & 
   \multicolumn{1}{c}{V367 Sct} & \multicolumn{1}{c}{CO Aur}\\
\noalign{\smallskip}
\hline
\noalign{\smallskip}
R$_{21}$(\fu)& 298\t12 & 104\t19 & 156\t14 & 175\t6 & 164\t12 & 153\t6 & 179\t6\\
R$_{21}$(\fd)& 103\t26 &  77\t7  &  46\t23 & 45\t9 & 24\t9 & 120\t9 \\
\noalign{\smallskip}
2\fu    &  418\t6 & 417\t9 &  433\t10& 421\t5 & 440\t7   & 450\t7 & 409\t8 \\
\fp     &  427\t7 & 436\t5 & 453\t10 & 450\t4 & 469\t8   & 402\t9 & 452\t24 \\
\fm     &  458\t7 & 466\t9 & 428\t14 & 456\t6 & 469\t10  & 499\t11& 473\t28 \\
2\fd    &  433\t13& 456\t10 &470\t49 & 523\t21& 300\t34  & 362\t12       \\
\noalign{\smallskip}
3\fu    &  226\t13&          &       & 230\t19& 255\t53  & 218\t36&183\t29 \\
2\fu+\fd&  224\t17& 283\t18 &        & 234\t13& 234\t33  \\
\fu+2\fd &  227\t34 \\      
2\fd--\fu&        &        &  165\t25\\
\noalign{\smallskip}
4\fu    &  \\
3\fu+\fd&  638\t30 \\
%2\fu+2\fd& \\
%3\fu--\fd&   \\
\noalign{\smallskip}
\hline
\hline

\end{tabular}
\end{flushleft}
\end{table*} 

Table 1 lists the $G_{i,j}$ values for each star's $f_{i,j}$'s, as
calculated on the basis of the phase values listed in Tab. 3, 4 and 5 of Paper
I. For each value the maximum error is also reported, as obtained from the
formal errors reported in Paper I.
We remind the reader that the real significance  of a component
was established star by star and hence we are dealing here with very reliable
values. Table 1 supplies the first synthetic description of the 
DMC light curves, as obtained from the available photometry.
\subsection{Separations between different order terms}
It is quite interesting to plot the $G_{i,j}$ values against their fit order.
Light curves of DMCs are often quoted as an example of erratic behaviour and
cycle--to--cycle variations, both in amplitude and in phase. In Paper I
we already proved that these light curve seem to be much more stable than
reported and that a frequency locked fit  yields a satisfactory
representation. Only in the cases of U TrA and EW Sct we found some slight
evidence of frequency or amplitude variations.

The suspicion that the DMC light curves have a
predictable behaviour is confirmed by the natural upper and lower limits
that can be easily observed in Fig. 1.
The second order terms are confined in the region just below 3/2$\pi$;
the third order terms have $\pi/2 < G_{i,j} < \pi$, the fourth order ones
cluster around 2$\pi$ (or 0), the fifth order ones seem to have
$\pi < G_{i,j} < 3/2 \pi$.

The mean $G_{i,j}$ values are 4.30\t0.34 rad for the second order (i.e.
$\mid i\mid + \mid j\mid$=2),
2.20\t0.23 rad for the third one, 6.24\t0.31 for the fourth one, 3.85\t0.21
for the fifth one. These mean values are roughly equispaced, with a slight
tendency to increase: indeed, the differences between the mean $G_{i,j}$ of adjacent
orders are 2.10, 2.24, 2.39 rad, respectively. The latter result and the
boundary values  established above yield an experimental confirmation of the
conjectures first expressed by Antonello (1994b) about the extension to DMCs
of the rule of uniformity of phase differences in monoperiodic Cepheids.
However, the observed separation ($\sim$2.2 rad) is a bit larger than expected
($\pi$/2) in the case of adiabatic pulsations in  a one--zone model.
\begin{figure}
%\picplace{9.0truecm}
%\epsfxsize=8.5cm
%\epsffile{fig1.pap2}
\psfig{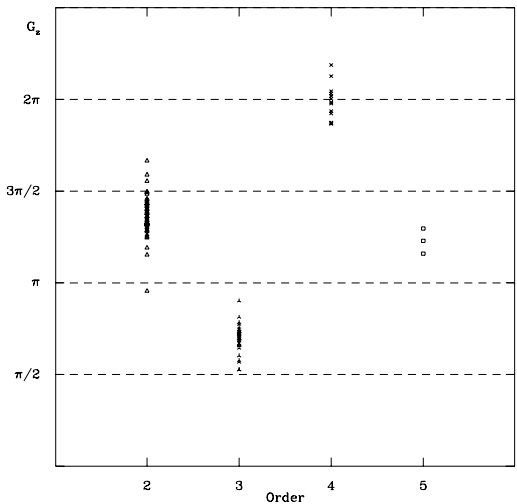}
\caption[]{A well defined separation is obtained when plotting the generalized
phase values $G_{i,j}$ in function of the respective order}
\end{figure}
\begin{figure}
%\epsfxsize=7.0truecm
%\epsffile[100 180 102 185]{fig2.pap2}
%\picplace{13.7truecm}
%\psfig{file=fig2.pap2,height=8truecm,width=7truecm}
\psfig{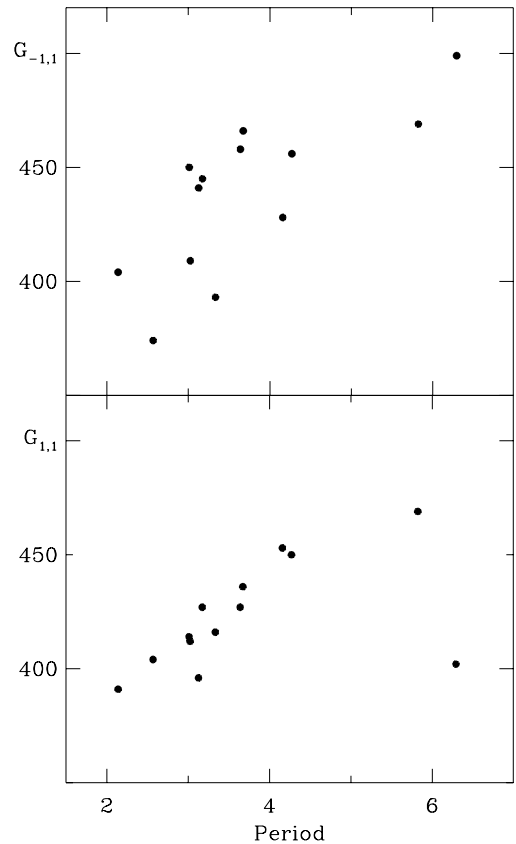}
\caption[]{The cross--coupling terms \fp and \fm are present in all the 
DMC light curves. While the $G_{-1,1}$  progression does not show any
particularity, the $G_{1,1}$  progression seems to be interrupted at 
$P\simeq$6.0 d}
 \end{figure}
\begin{figure} 
%\picplace{9.7truecm}
%\epsfxsize=8.5cm
%\epsffile{fig3.pap2}
\psfig{file=fig3.pap2,height=10truecm}
\caption[]{The amplitude of the cross--coupling terms \fp  decreases
towards longer periods and it reaches its minimum at P$\sim$6.0 d (lower
panel); the amplitude of the \fm term is small over the whole range of 
period values}
 \end{figure}
\subsection{The $G_{i,j}$ progressions}
The second order $G_{i,j}$ values (2\fu, \fp, \fm, 2\fd) range from
3.00 to 5.23 rad; it was expected to see a little 
spread of the $G_{0,2}$ values owing to the resonance at 3.0 d.
Indeed the two extrema are just related to
the 2\fd components  of the BQ Ser (the DMC approaching resonance from
the shorter periods) and EW Sct (the DMC approaching resonance from
the longer periods) light curves. Antonello (1994a) reported
another possible resonance between the third overtone and the \fp term 
near 6.5 d.
Later, Antonello (1994b) discussed a preliminary progression of the $G_{1,1}$ 
values and stressed the importance of verifying the
position of the points related to V367 Sct and EW Sct light curves.
This can now be carefully done as Fig. 2.
In the lower panel the last point (4.02\t0.09 rad, V367 Sct, $P$=6.293 d) is
clearly out of the  progression followed by the other points (the last is
at 4.69\t0.08 rad, EW Sct, $P$= 5.823 d). Moreover, the progressive weakening
of the amplitude of the \fp term is clearly visible in  Fig. 3 (lower panel);
the same trend can be evidenced by considering different types of normalized
amplitudes and mode energies and it must be considered as a well established
fact. On the other hand, if
we look at the \fm term we observed a smoother behaviour considering both   
the $G_{-1,1}$ progression (Fig. 2, upper panel) and 
its amplitude (Fig. 3, upper panel).
%Also the 3\fu value for V367 Sct is slightly lower than expected one.
All these facts  strongly supporting the action of a
resonance effect involving the \fp term.

The third order progressions  ($G_{2,1},
G_{1,2}, G_{-1,2}$) do not show any particular feature except a slight
tendency of increasing values for longer periods. The fourth order
term $G_{3,1}$ progression  mimics, at a different level, the $G_{3,0}$
one.
\section{Double--mode, classical and $s$--Cepheids}
\begin{figure*} 
%\picplace{18.0truecm}
%\epsfxsize=18.5cm
%\epsffile{fig4.pap2}
\psfig{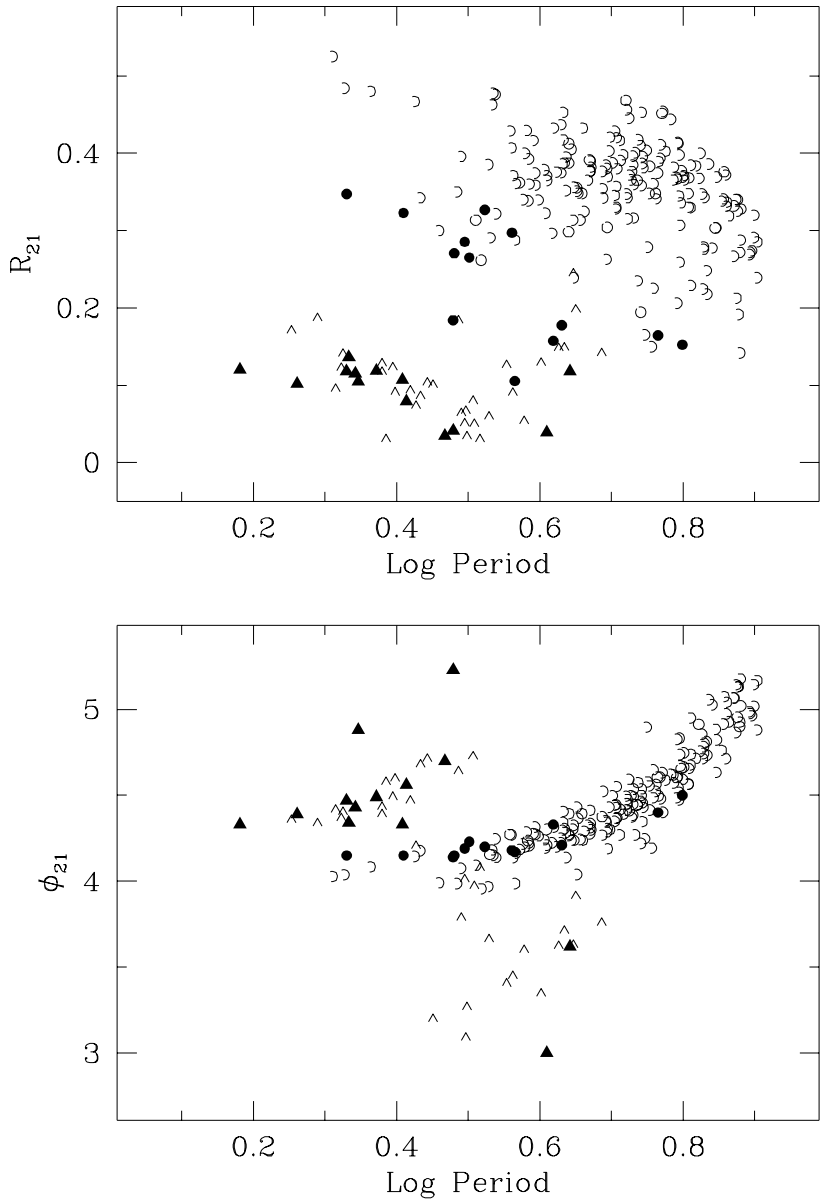}
\caption[]{The Fourier parameters $\phi_{21}$ and $R_{21}$ provide a powerful
discrimination between $F$ and 1$O$ pulsators. 
Dots: single--mode Classical
Cepheids. Triangles: $s$--Cepheids. Filled dots: Fundamental radial mode
of DMCs. Filled triangles: 1$O$ radial mode of DMCs}
 \end{figure*}
Figure 3 represents the actual scenario of single-- and double--mode Cepheids
with $P<$8 d in our galaxy.
The $\phi_{21}$--$P$ plane provides a quantitative 
representation of the  Hertzsprung progression, formed by the
Classical
Cepheids (open circles); the $\phi_{2,1}$ points obtained from the $F$ mode
 of DMCs (filled circles)
are perfectly superimposed. The ``$Z$" progression is defined by the 
single--mode Cepheids which do not follow the Hertzsprung progression; 
in our opinion, these stars are to be considered the true $s$--Cepheids and
their 1$O$ nature is now 
definitely established by the perfect superimposition of the $\phi_{21}$
values obtained from the 1$O$ mode of DMCs.

There is again a some underlying confusion about the nomenclature of these variables.
The general assumption is that ``$s$--Cepheids
have a quasi--sinusoidal light curve", but to prove it 
it is necessary to perform a quantitative analysis: in such a case
it is easy to establish the quasi--sinusoidal shape.
Other criteria (first look evaluation, full amplitude values, ....) are quite
arbitrary and not as useful as
the amplitude ratios $R_{21}$ for \fu (i.e. the ratio between
the amplitude of \fu and its harmonic 2\fu) and \fd (between the amplitude of
\fd  and 2\fd). As the lower panel of Fig. 3 shows, the stars 
forming the ``$Z$" sequence also show a small $R_{21}$ value and hence
the light curve deviates very slightly from a sinewave shape.
However, it should be noted that the $R_{21}$ values for the $F$--mode of
AX Vel (particularly), GZ Car, BQ Ser are smaller than the expected ones;
also the $R_{21}$ value
of the $F$--mode of VX Pup is small and it is located in the gap between
the two sequences, very close to that of the single--mode AV Cir. 
Hence, the Fourier parameters have to be considered globally
to perform a reliable mode identification.

The $G_{i,j}$ values of the fit of the CO Aur data were not used in the
 previous discussion because they strongly deviate from the progressions described
by the others; this is undoubtedly due to the 1$O$/2$O$ pulsating nature of
this star. It should be noted that the $G_{i,j}$ values are deviating only when
considering the period, but they are in the range of those of $F/$1$O$
pulsators (see Fig. 2 for the \fp and \fm cases).
In Paper I we already discussed its perfectly sine--shaped 2$O$ light curve.
\section{Conclusions}
In the previous sections we supplied a quantitative description of the
DMC light curves by means of their Fourier parameters and their progressions
as a  function of the period. We detected and
determined very precise ranges of generalized phase differences in
function of the order fit, also establishing the uniformity of the Fourier
parameter distribution. In such a context, these uniformities make the DMC
light curves very similar to the single--mode Cepheid ones. Hence, the physical
properties of the pulsation should be the same and, at least to a first
approximation, the adiabatic approximation provides a satisfactory
understanding of the phenomenon (Antonello 1994b)

It will be very interesting now to verify the properties of the generalized
phase differences in the more numerous sample constituted by the DMCs of the
Large Magellanic Clouds (Alcock et al. 1996, Welch et al. 1996); to do this,
we consider it an essential step to apply the frequency analysis described in
Paper I in order to avoid any spurious or unreliable result.
\begin{acknowledgements} The authors wish to thank E. Antonello for useful
discussions and  J. Vialle for the improvement of the English form of the
manuscript.
\end{acknowledgements}

\end{document}